\documentclass[aps,prx,twocolumn,superscriptaddress,longbibliography]{revtex4-1}

\usepackage[english]{babel} 
\usepackage[latin1]{inputenc}
\usepackage[usenames,dvipsnames]{color}
\usepackage{graphicx}
\usepackage{amsmath,amssymb,amsfonts}
\usepackage[colorlinks=true,linkcolor=blue,urlcolor=blue,citecolor=blue]{hyperref}
\usepackage[percent]{overpic}

\newcommand {\mc} {\mathcal}
\newcommand {\lan} {\left \langle}
\newcommand {\ran} {\right \rangle}

\newcommand {\sign} {\mathrm{sign}}

\begin{document}

\title{Crystallization of Levitons in the fractional quantum Hall regime}

\author{Flavio Ronetti}
\email{ronetti@fisica.unige.it}
\affiliation{Dipartimento di Fisica, Universit\`a di Genova, Via Dodecaneso 33, 16146, Genova, Italy}
\affiliation{CNR-SPIN, Via Dodecaneso 33, 16146, Genova, Italy}
\affiliation{Aix Marseille Univ, Universit\'e de Toulon, CNRS, CPT, Marseille, France}
\author{Luca Vannucci}
\affiliation{Dipartimento di Fisica, Universit\`a di Genova, Via Dodecaneso 33, 16146, Genova, Italy}
\affiliation{CNR-SPIN, Via Dodecaneso 33, 16146, Genova, Italy}
\author{Dario Ferraro}
\affiliation{Istituto Italiano di Tecnologia, Graphene Labs, Via Morego 30, I-16163 Genova, Italy}
\author{Thibaut Jonckheere}
\affiliation{Aix Marseille Univ, Universit\'e de Toulon, CNRS, CPT, Marseille, France}
\author{J\'er\^ome Rech}
\affiliation{Aix Marseille Univ, Universit\'e de Toulon, CNRS, CPT, Marseille, France}
\author{Thierry Martin}
\affiliation{Aix Marseille Univ, Universit\'e de Toulon, CNRS, CPT, Marseille, France}
\author{Maura Sassetti}
\affiliation{Dipartimento di Fisica, Universit\`a di Genova, Via Dodecaneso 33, 16146, Genova, Italy}
\affiliation{CNR-SPIN, Via Dodecaneso 33, 16146, Genova, Italy}

\begin{abstract}
Using a periodic train of Lorentzian voltage pulses, which generates soliton-like electronic excitations called Levitons, we investigate the charge density backscattered off a quantum point contact in the fractional quantum Hall regime. We find a regular pattern of peaks and valleys, reminiscent of analogous self-organization recently observed for optical solitons in non-linear environments. This crystallization phenomenon is confirmed by additional side dips in the Hong-Ou-Mandel noise, a feature that can be observed in nowadays electron quantum optics experiments.  
\end{abstract}

\maketitle

\section{Introduction}
The emergence of self-organized ordered patterns is a wide and fascinating field of physics, including, among its most intriguing examples, the formation of structures repeating themselves with a regular shape in time. In this context, observations of optical solitons in a non-linear background propagating with a spontaneously ordered temporal profile have been recently reported \cite{Cole17,Herink17,Krupa17}. In the framework of electron quantum optics \cite{Grenier11b, Bocquillon14,Bauerle18}, a train of Lorentzian voltage pulses naturally emerges as the best candidate to realize the solid state analogue of optical solitons, namely robust ballistically propagating wave-packets carrying a single electron with no additional particle-hole pairs \cite{levitov96, keeling06, Rech16, Vannucci17,Glattli18}. These minimal excitations, called \emph{Levitons}, represent one of the most reliable tools to inject single electronic states into ballistic channels of meso-scale devices \cite{dubois13, grenier13, Misiorny18, Ferraro15}, and have been recently exploited to reproduce some famous quantum-optical experiments, such as Hanbury-Brown-Twiss (HBT) or Hong-Ou-Mandel (HOM) interferometry, at the fermionic level \cite{dubois13-nature, Jullien14}. These fascinating experimental results open up the possibility of exploiting Levitons as flying qubits with appealing applications for quantum information processing \cite{Glattli16_physE,Glattli18}. Moreover, similarly to solitons, $q$ different Levitons travel unhindered along one-dimensional electronic edge states and can be controllably superimposed, thus forming many-body states called multi-electron Levitons or, simply, $q$-Levitons \cite{Glattli16_physE,Moskalets18}.\\
However, it is well known that one dimensional electronic systems are drastically affected by electron-electron interactions. The latter can induce, for instance, the arrangement of electrons in a static regular pattern in space, a phenomenon known as Wigner crystallization \cite{Wigner34,Schulz93,Deshpande08,Wang12,Pecker13,Traverso13,Deshpande10,Matveev04}.
A seminal example of strongly interacting electron systems is provided by the fractional quantum Hall (FQH) effect \cite{Stormer99}. Here, one dimensional channels at the boundaries of the Hall bar are described in terms of chiral Luttinger liquids \cite{Wen95}, whose direction of propagation is imposed by the external magnetic field. In these systems the connection between time and space given by chirality opens the way to the possible realization of the real-time version of the interaction-induced crystallization by applying time dependent voltage pulses directly to the edge channels.\\
In this paper, we propose FQH states belonging to the Laughlin sequence \cite{Laughlin83}, where a single mode exists on each edge, as a testbed to observe the crystallization of robust $q$-Leviton excitations in condensed matter systems. Here, the charge density reflected by a quantum point contact (QPC) shows a $q$-peaked structure as a consequence of the interaction-induced rearrangement in the time domain, in open contrast to the featureless profile observed in the integer case. To confirm the correlated character of the crystal state, we demonstrate that these features generate unexpected side dips in the noise profile of HOM collisional experiments, which are within reach for the nowadays technology \cite{Ol'khovskaya08,Bocquillon13, dubois13-nature, Wahl14, Freulon15, Marguerite16}.\\ The paper is organized as follows. Sec. \ref{sec:Model} exposes the model and the setup. In Sec. \ref{sec:density}, we present the derivation of the excess density and we discuss the crystallization of Levitons. Then, in Sec. \ref{sec:experimental}, we describe possible experimental signatures of the crystallization of Levitons in the HOM setup. Finally, Sec. \ref{sec:Conclusions} is devoted to our conclusion.

%
%
\section{Model \label{sec:Model}} We consider a 4-terminal FQH bar in the presence of a QPC, as shown in the inset of Fig.\ \ref{fig:setup_hom}. The Hamiltonian $H = H_{\rm 0} + H_{\rm s} + H_{\rm T}$ consists of edge states, source and tunneling terms respectively. For a quantum Hall system with filling factor $\nu$ in the Laughlin sequence $\nu=1/(2n+1)$ \cite{Laughlin83}, with $n \in \mathbb N$, a single chiral mode emerges at each edge of the sample. The effective Hamiltonian for the edge states reads ($\hbar=1$) \cite{Wen95}
\begin{equation}
	\label{eq:H_wg}
	H_{\rm 0} = \sum_{r=R,L} \frac{v}{4\pi} \int dx \left[ \partial_x \Phi_r(x) \right]^2.
\end{equation}
Here, bosonic fields $\Phi_{R/L}$, satisfying $[\Phi_{R/L}(x), \Phi_{R/L}(y)] = \pm i \pi \sign(x-y)$, describe right and left moving excitations propagating at velocity $v$ along the edge.
Annihilation fields for Laughlin quasiparticles carrying fractional charge $-\nu e$ (with $e>0$) are defined through the standard procedure of bosonization \cite{Wen95}. They read 
\begin{equation}
\psi_{R/L}(x) = \frac{\mathcal{F}_{R/L}}{\sqrt{2\pi a}} e^{-i \sqrt \nu \Phi_{R/L}(x)},
\end{equation}
where $a$ is a short-distance cut-off and $\mathcal{F}_{R/L}$ are the Klein factors \cite{Wen95,vonDelft98,Martin05}.
The source term
\begin{equation}
\label{H_s}
	H_{\rm s} = \sum_{r=R,L} \int dx \, \Theta(\mp x-d) V_r(t) \rho_r(x)
\end{equation}
couples charge densities $\rho_{R/L} (x) = \pm \frac{e \sqrt{\nu}}{2\pi} \partial_x \Phi_{R/L}(x)$ with two voltage gates acting separately on the right and left moving excitations. Here, the step function $\Theta(\mp x-d)$ describes the experimentally relevant situation of infinite, homogeneous contacts. Equations of motion for the bosonic fields in the presence of the source term are solved in terms of the single-variable fields $\phi_{R/L}$ in the equilibrium configuration $V_L=V_R=0$. Solutions read (see Appendix \ref{app:equation})
\begin{equation}
\label{eq:eq_motion}
	\Phi_{R/L}(x,t) = \phi_{R/L}\left(t \mp \frac x v \right) - e \sqrt \nu \int_0^{t \mp \frac x v} dt' V_{R/L}(t').
\end{equation}
This characteristic chiral dynamics is a consequence of the linear dispersion of edge states for all filling factors in the Laughlin sequence.\\
The soliton crystal phase in the FQH regime arises at its best when considering purely electronic excitations devoid of additional particle-hole pairs, i.e.\ the aforementioned Levitons \cite{dubois13-nature}. As both theory and experiments indicate, such unique states emerge in response to well defined voltage pulses of Lorentzian shape \cite{levitov96,keeling06,dubois13-nature,Rech16}. To make contact with experiments, we will thus consider a periodic train of Lorentzian pulses
\begin{equation}
V(t)=\sum\limits_{k=-\infty}^{+\infty}\frac{V_0}{\pi}\frac{W^2}{W^2+(t-k \mathcal{T})^2},
\end{equation}
with period $\mathcal{T}=\frac{2\pi}{\omega}$, amplitude $V_0$ and width $2W$. In particular, we will focus on quantized pulses carrying an integer charge $-qe=\frac{e^2\nu}{2\pi}\int_{0}^{\mathcal{T}}dt V(t)$, here named $q$-Levitons.\\
Finally, the tunneling between the two edges occurs through a QPC at $x=0$. Assuming that the QPC is working in the weak backscattering regime, the tunneling of Laughlin quasiparticles between opposite edges is the only relevant process \cite{Kane94,Saminadayar97,dePicciotto97,Ferraro10_PRB}. The corresponding Hamiltonian is $H_{\rm T} = \Lambda \psi^\dag_R(0) \psi_L(0) + \mathrm{h.c.}$, with $\Lambda$ the constant tunneling amplitude.

\section{Density and Leviton crystallization \label{sec:density}} 
The formation of a $q$-Leviton crystal can be seen from the behavior of the excess charge density, defined as
\begin{equation}
\label{eq:exc_dens}
\Delta \rho_{R/L}(x,t)=\langle \rho_{R/L}(x,t)\rangle-\langle \rho^{(0)}_{R/L}(x,t)\rangle .
\end{equation}
Here, density operators evolve in time according to Eq.\ \eqref{eq:eq_motion}, and 
\begin{equation}
	\label{eq:J0}
	\rho^{(0)}_{R/L}(x,t) 
	= \pm \frac{e \sqrt \nu}{2\pi} \partial_x \phi_{R/L}(t_\mp)
\end{equation}
is the charge density operator at equilibrium ($V_R=V_L=0$). The assumption of weak backscattering regime allows us to calculate the excess charge density perturbatively in the tunneling Hamiltonian $H_T$. Thermal averages are thus performed over the initial equilibrium density matrix in the absence of tunneling.\\
Calculations are usefully carried out in terms of quasiparticle correlation functions $G^{(qp)}_{R/L}(x',t';x,t) = \langle \psi^\dagger_{R/L}(x',t')\psi_{R/L}(x,t)\rangle$.
The equilibrium quasiparticle correlation functions can be evaluated through standard bosonization technique and yield $G^{(0)}_{R/L}(x',t';x,t) = G_0 (t'_\mp-t_\mp)$, with (we use the notation $t_\mp=t \mp \frac x v$ throughout the paper)
\begin{equation}
G_0(\tau) = \frac{1}{2\pi a} \left[ \frac{\pi k_{\rm B} \theta \tau}{\sinh\left( \pi k_{\rm B} \theta \tau \right) \left(1+i \omega_c \tau \right)}\right]^{\nu}.
\end{equation}
Here, $\theta$ is the temperature and $\omega_c=\frac{v}{a}$ is the high-energy cutoff. The fact that $G^{(0)}_{R/L}$ effectively depends on a single-variable function is a joint consequence of the chirality of Laughlin states and translational invariance at equilibrium.
Deviations from the equilibrium correlators, defined as $\Delta G^{(qp)}_{R/L}=G^{(qp)}_{R/L}-G^{(0)}_{R/L}$, carry all the information about the propagation of Levitons. \\ 
Let us initially create only right-moving excitations, by imposing $V_R(t)=V(t)$ and $V_L(t)=0$. This experimental configuration, in which pulses from a single source are partitioned against a beam splitter, is usually termed HBT setup \cite{Brown56,dubois13, Bocquillon12}. In this configuration the right-moving excess correlator reads
\begin{align}
	& \Delta G^{(qp)}_R(x',t';x,t) =-2 iG_0 (t'_{-} - t_{-}) \times \nonumber \\
	& \quad \times \sin \left( \frac{\pi(t'_{-} - t_{-})}{\mathcal{T}} \right) \sum\limits_{k=1}^{q}\varphi_k(t_{-})\varphi_k^{*}(t'_{-}),
	\label{eq:DeltaG}
\end{align}
where the functions
\begin{equation}
\label{eq:phi}
\varphi_k(t)=\sqrt{\frac{\sinh\left(2\pi \frac{W}{\mathcal{T}}\right)}{2}}\frac{\sin^{k-1}\left(\pi\frac{t-i W}{\mathcal{T}}\right)}{\sin^{k}\left(\pi\frac{t+i W}{\mathcal{T}}\right)}
\end{equation}
are periodic wave functions with period $2\mc T$ \cite{Glattli16_physE,glattli2016method}. They generalize the set of single-electron wave functions introduced for the Lorentzian pulse \cite{grenier13,Ferraro13,Moskalets15}, and form a complete orthonormal basis, thus satisfying the condition $\int_{0}^{\mathcal{T}}\frac{dt}{\mathcal{T}}\varphi_k(t)\varphi^{*}_{k'}(t)=\delta_{k,k'}$. Let us notice that Eq.\ \eqref{eq:DeltaG} reduces to the so called single-electron coherence function (a crucial tool in the context of electron quantum optics) in the limit of free fermions ($\nu=1$) and infinite period \cite{grenier13,Ferraro13}.
Excess correlators for quasiholes can be defined similarly as $\Delta G^{(qh)}_R = \langle \psi_R(x',t')\psi_R^{\dagger}(x,t)\rangle-G^{(0)}_R$,
yielding 
\begin{equation}
\Delta G^{(qh)}_R = 2 i G_0(t'_{-}-t_{-})\sin\left(\pi \frac{t_{-}'-t_{-}}{\mathcal{T}}\right)\sum\limits_{k=1}^{q}\varphi_{k}^{*}(t_{-})\varphi_{k}(t_{-}').
\end{equation}
The excess density in Eq. \eqref{eq:exc_dens} varies significantly if evaluated \textit{before} or \textit{after} the scattering of injected particles at the QPC. Indeed, before the scattering we have $\Delta \rho_L(x,t)$=0, while $\Delta \rho_{R}(x,t)$ can be readily obtained by evaluating the excess quasiparticle correlator at equal times and positions. In the region $-d<x<0$ (that is, downstream of the contact but still before the QPC) we find
\begin{equation}
\label{eq:rho0}
\Delta \rho_{R}(x,t)=\frac{e }{v\mathcal{T}}\sum\limits_{k=1}^{q}\left|\varphi_k (t_{-})\right|^2=\frac{e^2 \nu}{2\pi v}V (t_{-}),
\end{equation}
since $\left|\varphi_k(t)\right|^2=\frac{e\nu V(t)}{q\omega}$ for each $k$. We note that Eq. \eqref{eq:rho0} is nothing but the single-particle density of a $q$-particle state described by a Slater determinant formed by the set of wave functions $\{\varphi_k\}, k=1,...,q$ \cite{grenier13}. Remarkably, this excess density does not display any qualitative difference between the integer and the fractional case.

\begin{figure}
	\centering
	\includegraphics[width=1\linewidth]{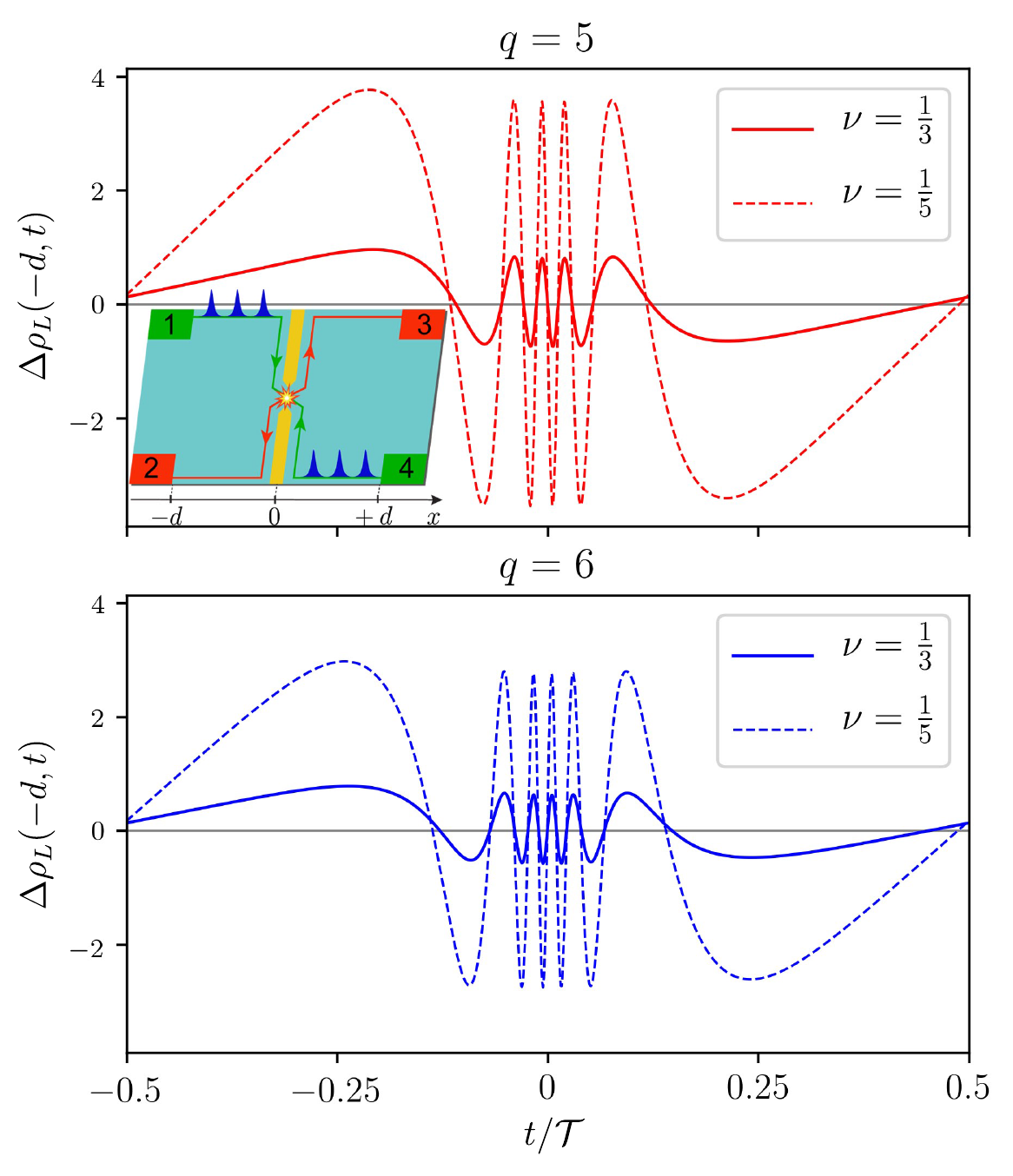}
	\caption{(Color online) Inset: sketch of the setup. The pulses originate from contacts 1 (top edge, $x \leq -d$) and 4 (bottom edge, $x \geq +d$) and propagate along the edge states of a FQH system. They may be either reflected or transmitted at $x=0$, due to the presence of a QPC. Main panels: Excess charge density $\Delta \rho_{L}(-d,t)$ evaluated in terminal $2$ (i.e. $x=-d$), in the presence of a single source for $q=5$ and $q=6$, in units of $\frac{e \left|\Lambda\right|^2\omega_c}{2\pi v^3}$. Two different filling factors are considered: $\nu=\frac{1}{3}$ (solid lines) and $\nu=\frac{1}{5}$ (dashed lines). The other parameters are $W=0.04 \mathcal{T}$, $k_{\rm B} \theta=10^{-3}\omega$ and $\omega=0.01 \omega_c$. }
	\label{fig:setup_hom}
\end{figure}

\noindent Non-linear tunneling, typical of the interacting FQH phase, is however expected to influence the propagation of Levitons \emph{after} the scattering at the QPC \cite{Kane94,Kane92}. We thus focus on the excess density backscattered into the left-moving channel, namely $\Delta \rho_L(x,t)$, with $x<0$. Since the QPC is assumed to work in the weak backscattering regime, we are allowed to set up a perturbative expansion in the tunneling amplitude $\Lambda$ for the charge density operator $\rho_{L}(x)=-\frac{e \sqrt \nu}{2\pi}\partial_x\Phi_{L}(x)$, which reads
\begin{align}
	\rho_{L}(x,t) &=-\frac{e^2 \nu}{2\pi v} V_{L} \left(t \mp \frac x v\right)+ \rho^{(0)}_{L}(x,t) + \rho^{(1)}_{L}(x,t)\nonumber+ \\&+ \rho^{(2)}_{L}(x,t) + o(\Lambda^3). 
\end{align}
Here, $\rho^{(0)}_{L}(x,t)$ is given by Eq.\ \eqref{eq:J0}, while subsequent contributions are given by
\begin{align}
	& \rho^{(1)}_{L}(x,t) = i \int_{-\infty}^t dt' \left[ H_{\rm T}(t'), \rho^{(0)}_{L}(x,t) \right]\nonumber\\& = - \Theta(- x) i \nu e \left\{ \frac{\Lambda}{v} \psi^{\dagger}_{R} \left(0,t_+ \right)\psi_{L} \left(0,t_+ \right)- \mathrm{h.c.} \right\}, \\
	\label{eq:J2}
	& \rho^{(2)}_{L}(x,t) = \nonumber \\
	& = i^2 \int_{-\infty}^t dt' \int_{-\infty}^{t'} dt'' \left[  H_{\rm T}(t'') , \left[ H_{\rm T}(t'), \rho^{(0)}_{L}(x,t) \right] \right]   \nonumber \\
	& = i \int_{-\infty}^{t_+} dt'' \left[ \Lambda\psi^{\dagger}_R(0,t'')\psi_L(0,t'') + \mathrm{h.c.} \, , \rho^{(1)}_{L}(x,t) \right].
\end{align}
Note that the step function in $\rho^{(1)}_{L}(x,t)$ is directly related to the effect of backscattering at $x=0$.\\
We thus get the excess charge density to lowest non-vanishing order in the tunneling, which reads
\begin{align}
	& \Delta\rho_{L}(x,t) = -\frac{e\nu|\Lambda|^2}{v} \int_{-\infty}^{t_+} dt' \Big[\Delta G_R^{(qp)} (0,t';0,t_+) + \nonumber \\
	& \quad - \Delta G_R^{(qh)} (0,t';0,t_+)\Big] G_0 \left( t'-t_+\right) +\text{h.c.},	\label{eq:rho_main}
\end{align}
According to the completeness of the set $\{\varphi_k\}$, the above result can be recast in the more compact and physically insightful form
\begin{equation}
\label{eq:rho2_main}
\Delta\rho_{L}(x,t)=\frac{e|\Lambda|^2}{v^3\mathcal{T}}\sum\limits_{k=1}^{q}\sum\limits_{p=1}^{+\infty}\Re[c_{k,p}\varphi_k(t_{+})\varphi^{*}_{p}(t_{+})],
\end{equation}
where coefficients $c_{k,p}$ depend on the temperature $\theta$ and the filling factor $\nu$. In terms of the overlap integrals $g_{kp}(\bar{t})=\int_{0}^{\mathcal{T}}\frac{dt}{\mathcal{T}}\varphi_k(t+\bar{t})\varphi^{*}_p(t)$, they are given by
\begin{equation}
	c_{k,p}=-\frac{16\pi\nu v^2}{ \omega}\int_{-\infty}^{0}dt'g^{*}_{kp}(t')\sin\left(\frac{\pi t'}{\mathcal{T}}\right)\Im\left[G^2_0(t')\right].
\end{equation}
In an ordinary metallic system ($\nu=1$), they reduce to $c_{k,p}=\delta_{k,p}$, so that Eq. \eqref{eq:rho2_main} becomes simply $\Delta\rho_{L}(x,t)=\frac{e|\Lambda|^2}{v^3\mathcal{T}} V(t_{+})$. Thus backscattered pulses at $\nu=1$ maintain the same Lorentzian shape as the injected ones. \\
Conversely, the excess density in a Laughlin FQH system departs strongly from the trivial metallic result, as we show in Fig. \ref{fig:setup_hom} for $\nu=\frac{1}{3}$ and $\nu=\frac{1}{5}$ and different values of $q$. Here we focus on the excess density measured in terminal 2, i.e. for $x=-d$ \cite{position}.
Due to the strongly correlated background, the $q$-Leviton state backscattered off the QPC is rearranged into an oscillatory pattern with a number of peaks exactly equal to $q$, regardless of any other parameter. The amplitude of the oscillations increases with decreasing filling factor (that is, for stronger correlations). These patterns suggest that scattering at the QPC creates a correlated structure of $q$ separated and co-moving Levitons. In analogy with other strongly correlated phases in condensed matter \cite{Traverso12,Deshpande08,Rontani04}, we interpret this structure as a \textit{crystallization} of the $q$-Leviton state. However, in contrast to Wigner crystallization, the arrangement induced by interaction does not show a static profile but rather a propagating one, thus leading to the emergence of a regular structure in time and not only in space. Due to the soliton-like nature of Levitons \cite{levitov96}, this process presents an intriguing analogy with the formation of optical soliton crystal in the presence of a non-linear environment \cite{Cole17}, albeit in a completely different context.

In passing, let us comment about the parity of excess density shown in Fig. \ref{fig:setup_hom}. In this light, it is useful to further manipulate Eq. \eqref{eq:rho_main} in such a way that
\begin{align}
	\label{eq:rho_real_im}
	\Delta\rho_{L}(x,t)&=\frac{e|\Lambda|^2}{v^3\mathcal{T}}\sum\limits_{k=1}^{q}\sum\limits_{p=1}^{+\infty}\Big\{\Re[c_{k,p}]\Re[\varphi_k(t_{+})\varphi^{*}_{p}(t_{+})]+\nonumber\\&-\Im[c_{k,p}]\Im[\varphi_k(t_{+})\varphi^{*}_{p}(t_{+})]\Big\}.
\end{align}
Here, $\Re[\varphi_k(\tau)\varphi^{*}_{p}(\tau)]$ and $\Im[\varphi_k(\tau)\varphi^{*}_{p}(\tau)]$ are, respectively, an even function and an odd function of $\tau$, since $\varphi_k(\tau)=-\varphi^{*}_k(-\tau)$ (see Eq. \eqref{eq:phi}). It is thus clear that the excess density has not a definite parity with respect to $t_{+} = t + \frac x v$ for a generic value of $\nu$, as both an even term and an odd component are present in Eq.\ \eqref{eq:rho_real_im}. In the non-interacting case ($\nu=1$), the coefficients $c_{k,p}$ are real-valued and the excess density reduces to an even function of $t_{+}$.

\section{Experimental signatures in current noise \label{sec:experimental}}

A direct observation of the oscillating density would require a real-time measurement of the backscattered current with extremely high temporal resolution. Moreover, this observation alone would not be the conclusive proof of the crystallization process.  In order to indubitably relate the oscillations of the density to the crystallization of Levitons, one has to further investigate the density-density or current-current correlators \cite{Wang12,Gambetta14}. The very special nature of the $q$-Leviton crystal, which is not confined to a finite spatial region, but rather moves rigidly along the edges, lets us envisage an experimental test based on the cross-correlations of two flying crystallized patterns. In this light, we propose to perform a much more feasible zero-frequency measurement of current noise in a HOM experimental setup \cite{Hong87,Bocquillon13,dubois13}. In this configuration, a second train of Levitons (identical to the first one) is generated in terminal $4$ and delayed by a tunable time shift $t_D$. We describe the HOM setup by setting $V_R(t)=V(t)$ and $V_L(t)=V(t+t_D)$. A genuine crystallization process is expected to manifest as oscillations in the current noise analyzed as a function of the delay $t_D$. As a side note, let us observe that intensity-intensity correlation measurements are analogously performed to probe the crystallization of solitons in the optical domain \cite{Cole17}.\\ We thus focus on the zero-frequency cross-correlation between terminals $2$ and $3$, defined as
\begin{align}
\label{eq:def_noise}
	\mc S_{23} & = v^2\int_{0}^{\mathcal{T}} \frac{dt}{\mathcal{T}} \int\limits_{-\infty}^{+\infty} d \tau \left[ \lan \rho_{R}(d,t+\tau) \rho_{L}(-d,t) \ran \right. \nonumber \\
	& \quad \left. - \lan \rho_{R}(d,t+\tau) \ran \lan \rho_{L}(-d,t) \ran \right] .
\end{align}
A standard procedure is to normalize the HOM signal with respect to the HBT one \cite{Jonckheere12,Bocquillon13}. We thus define the ratio
\begin{equation}
\label{ratio}
\mathcal{R}=\frac{\mathcal{S}_{23}^{\rm HOM}(t_D)-\mathcal{S}_{23}^{\rm vac}}{2\mathcal{S}^{\rm HBT}_{23}-2\mathcal{S}_{23}^{\rm vac}},
\end{equation}
where $\mathcal{S}_{23}^{\rm HOM}$ and $\mathcal{S}_{23}^{\rm HBT}$ are the cross-correlators measured respectively in the HOM and HBT configurations discussed above and read
\begin{widetext}
\begin{align}
	\mc S^{\rm HOM}_{23}
	& = (\nu e)^2 |\Lambda|^2 \int_{0}^{\mathcal{T}} \frac{dt}{\mathcal{T}} \int\limits_{-\infty}^{+\infty} d \tau \left[G^{(qp)}_R \left(0,t+\tau-\frac{d}{v};0,t-\frac{d}{v} \right) G^{(qh)}_L \left(0,t+\tau-\frac{d}{v};0,t-\frac{d}{v} \right) + \right. \nonumber \\\label{sup_eq:noise1}
	& \quad + \left. G^{(qh)}_R \left(0,t+\tau-\frac{d}{v};0,t-\frac{d}{v} \right) G^{(qp)}_L \left(0,t+\tau-\frac{d}{v};0,t-\frac{d}{v} \right) \right],\\	
	\mc S^{{\rm HBT}}_{23} 
	& = (\nu e)^2 |\Lambda|^2 \int_{0}^{\mathcal{T}} \frac{dt}{\mathcal{T}} \int\limits_{-\infty}^{+\infty} d \tau \left[G^{(qp)}_{R}(0,t+\tau;0,t)+G^{(qh)}_{R}(0,t+\tau;0,t)\right]G_0(\tau).	\label{sup_eq:noiseHBTR}
\end{align}
Note that we have isolated the desired signal by subtracting equilibrium fluctuations 
\begin{equation}
	\label{sup_eq:noisevac}
	\mc S^{\rm vac}_{23}  = 2 (\nu e)^2 |\Lambda|^2 \int\limits_{-\infty}^{+\infty} d \tau G^2_{0}(\tau),
\end{equation}
obtained with both sources off. As for the excess charge density, these quantities are evaluated to lowest order in the tunneling in terms of quasiparticle and quasihole correlators. The result reads \cite{delay}
\begin{equation}
\mathcal{R}=1+\frac{\int_{0}^{\mathcal{T}}dt\int\limits_{-\infty}^{+\infty}  dt' \left(\Delta G^{(qp)}_{R}(0,t';0,t)\Delta G^{(qh)}_{L}(0,t';0,t)+\Delta G^{(qp)}_{L}(0,t';0,t)\Delta G^{(qh)}_{R}(0,t';0,t)\right)}{2\int_{0}^{\mathcal{T}}dt\int\limits_{-\infty}^{+\infty}  dt' \left(\Delta G^{(qp)}_{R}(0,t';0,t)+\Delta G^{(qh)}_{R}(0,t';0,t)\right)G_0(t'-t)},
\end{equation}
\end{widetext}
where we have used the relation $\Delta G^{(qp)/(qh)}_L(0,t';0,t)=\Delta G^{(qp)/(qh)}_R(0,t'+t_D;0,t+t_D)$. We now note that the completeness of the orthonormal set of wave functions guarantees that $\varphi_k(t+t_D)=\sum\limits_{p=1}^{+\infty}g_{kp}(t_D)\varphi_p(t)$. By using this result, the ratio can be conveniently formulated in terms of overlap integrals
\begin{equation}
\label{eq:ratio}
\mathcal{R}=1-\frac{\sum\limits_{k,k'=1}^{q}\sum\limits_{p,p'=1}^{+\infty}\Re\left[w_{pp'}^{k}g_{k'p}(t_D)g^{*}_{k'p'}(t_D)\right]}{v_q},
\end{equation}
where the coefficients $w_{pp'}^{k}$ and $v_{q}$ encodes the dependence on interaction and temperature and are given by
\begin{align}
	w_{pp'}^{k}&=\int_{0}^{\mathcal{T}} \frac{dt}{\mathcal{T}} \int\limits_{-\infty}^{+\infty}d\tau\hspace{1mm}\varphi_k(t)\varphi^{*}_k(t+\tau)\times\nonumber\\&\times\varphi_{p}(t)\varphi^{*}_{p'}(t+\tau)\sin^2\left(\frac{\pi\tau}{\mathcal{T}}\right)G^2_0(\tau),\\
	v_{q}&=\sum\limits_{k=1}^{q}\int\limits_{-\infty}^{+\infty}d\tau \sin\left(\frac{\pi \tau}{\mathcal{T}}\right)g^{*}_{kk}(\tau)G_0^2(\tau).\label{sup_eq:ratio_fin}
\end{align}

\begin{figure}
	\centering
	\includegraphics[width=1\linewidth]{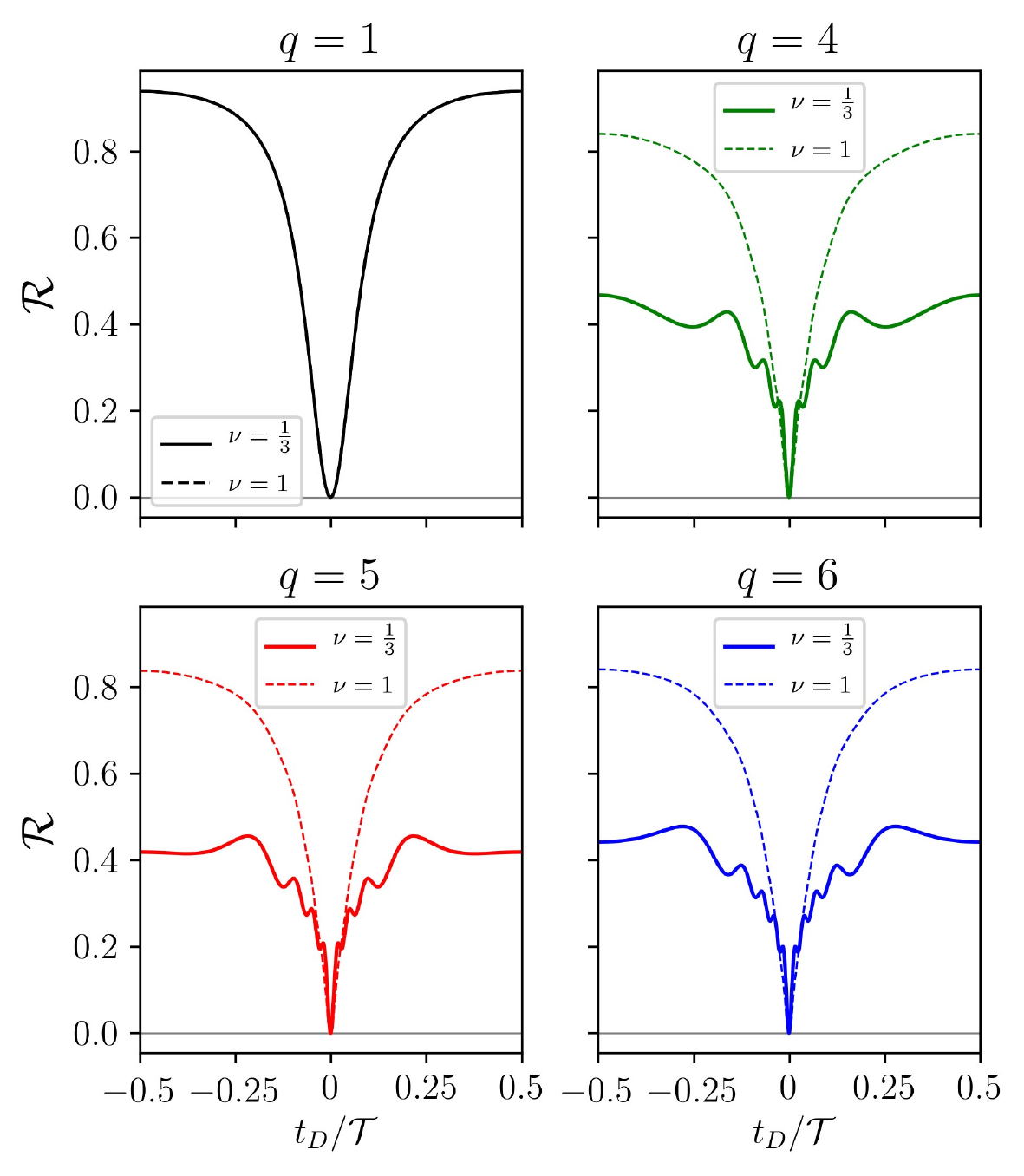}
	\caption{(Color online) Ratio $\mathcal{R}$ as a function of the time delay $t_D$ for $q=1$, $q=4$, $q=5$, $q=6$. The integer case (dashed lines) and the fractional case for $\nu=\frac{1}{3}$ (solid lines) are compared. The other parameters are $W=0.04 \mathcal{T}$, $k_{\rm B} \theta=10^{-3}\omega$ and $\omega=0.01 \omega_c$.}
	\label{fig:ratio}
\end{figure}

\noindent In the free fermion case and low temperature limit we find $w_{pp'}^{k}=\delta_{k,p}\delta_{k,p'}$ and $v_{q}=q$. Then, Eq. \eqref{eq:ratio} reduces to $\mathcal{R}=1-\frac{1}{q}\sum\limits_{k=1}^{q}\sum\limits_{k'=1}^{q}\left|g_{k'k}(t_D)\right|^2$, in accordance with previous results \cite{Jonckheere12,Glattli16_physE}.\\ The HOM ratio at $\nu=1$ consists of a single, smooth dip shown with dashed lines in Fig.\ \ref{fig:ratio} for different values of $q$. The absence of any additional structure at $\nu=1$ confirms the uncorrelated nature of Levitons in the Fermi-liquid state. Conversely, full lines in Fig. \ref{fig:ratio} show the behavior of $\mc R(t_D)$ at fractional filling $\nu=\frac 1 3$ for the same values of $q$.
We first notice that completely destructive interference between the two signals always occurs at $t_D=0$ (as demonstrated by the total central dip), whether the system is interacting or not. This shows that electron-electron interactions in single-edge-mode Laughlin states do not induce decoherence effects, in contrast with the role played by interactions in the $\nu=2$ integer quantum Hall effect, where two co-propagating edge states exist \cite{Ferraro14,Wahl14}.
At $q=1$, the ratio exhibits the same behavior for integer and fractional filling factors \cite{Rech16}. This is related to the fact that backscattering of a single Leviton generates a simple signal with no internal peak/valley structure. For higher values of $q$, rearrangement of $q$-Leviton excitations generates peculiar features that distinguish between the non-interacting and the strongly correlated phase. Plots at $\nu=\frac{1}{3}$ clearly show the presence of oscillations in the current-current correlators for $q>1$, with $2q-2$ new dips aside of the principal one at $t_D=0 $. It is interesting to notice that their arrangement bears similarities with the behavior of $\Delta \rho_L(x,t)$ shown in Fig.\ \ref{fig:setup_hom}. Indeed, as for the excess density, the spacing between maxima/minima of $\mc R(t_D)$ tends to widen while approaching the ends of the period.
These features unambiguously identify the effects of the strongly correlated FQH phase on Leviton excitations, in striking contrast with the uncorrelated Fermi liquid phase. A similar pattern was predicted in Ref. \cite{Wahl14} and experimentally observed in Ref. \cite{Freulon15}, where the internal peak/valley structure is generated by a fractionalization effect in a $\nu=2$ quantum Hall interferometer. Here we argue that the new side dips must be relatedto the unprecedently reported process of crystallization of $q$-Levitons in FQH edge states, as no fractionalization occurs in the single-edge-mode Laughlin sequence. Therefore, the appearance of local maxima and minima in the current-current correlators at fractional filling factors proves the existence of a $q$-Leviton crystal in the time domain induced by interactions.

\begin{figure}[h]
	\centering
	\includegraphics[width=1\linewidth]{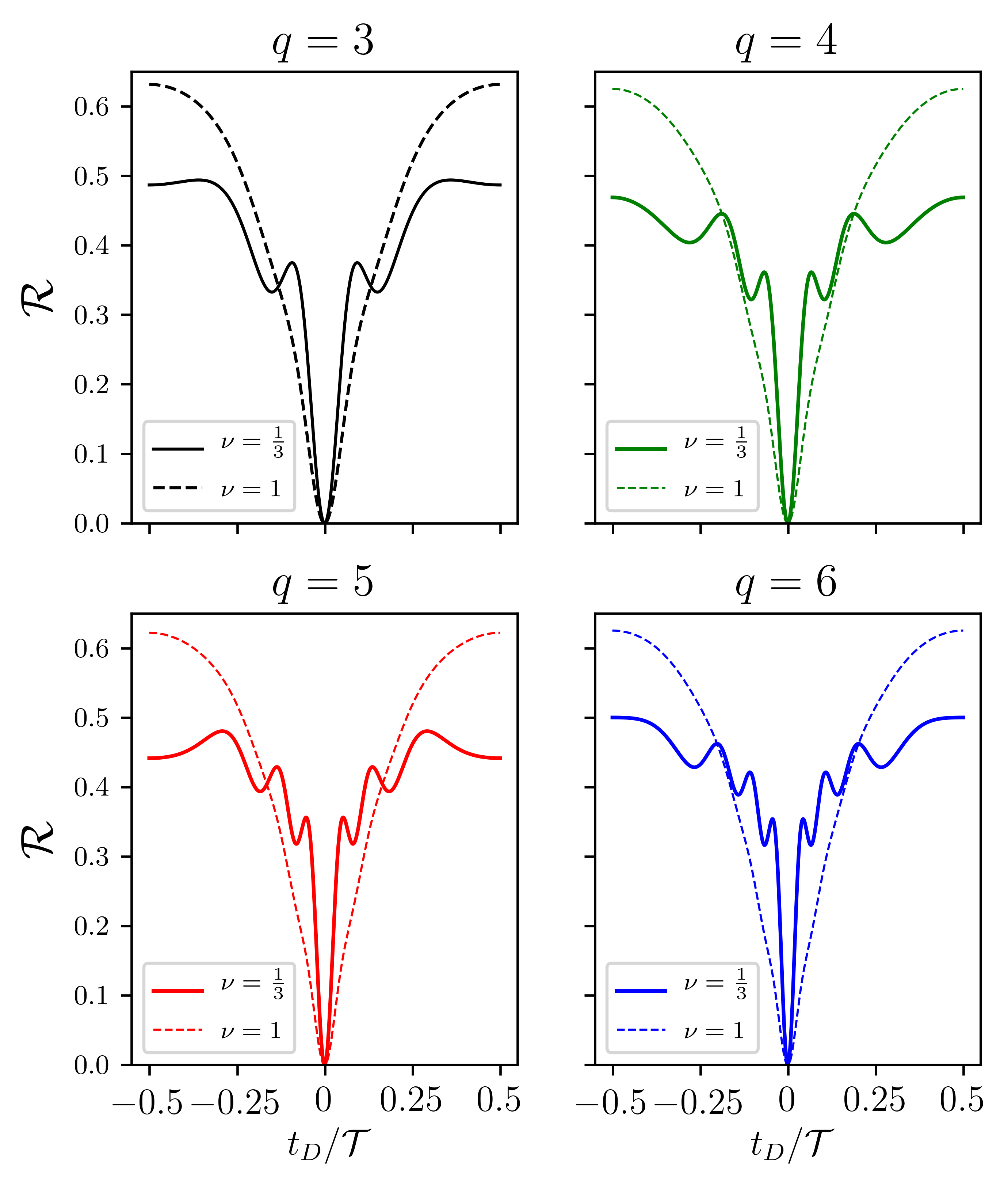}
	\caption{(Color online) Ratio $\mathcal{R}$ as a function of the time delay $t_D$ for $q=3$, $q=4$, $q=5$, $q=6$. The integer case (dashed lines) and the fractional case for $\nu=\frac{1}{3}$ (solid lines) are compared. The other parameters are $W= 0.1 \mc T$ and $\omega=0.01 \omega_c$.}
	\label{fig:ratio2}
\end{figure}

\begin{figure}[h]
	\centering
	\includegraphics[width=1\linewidth]{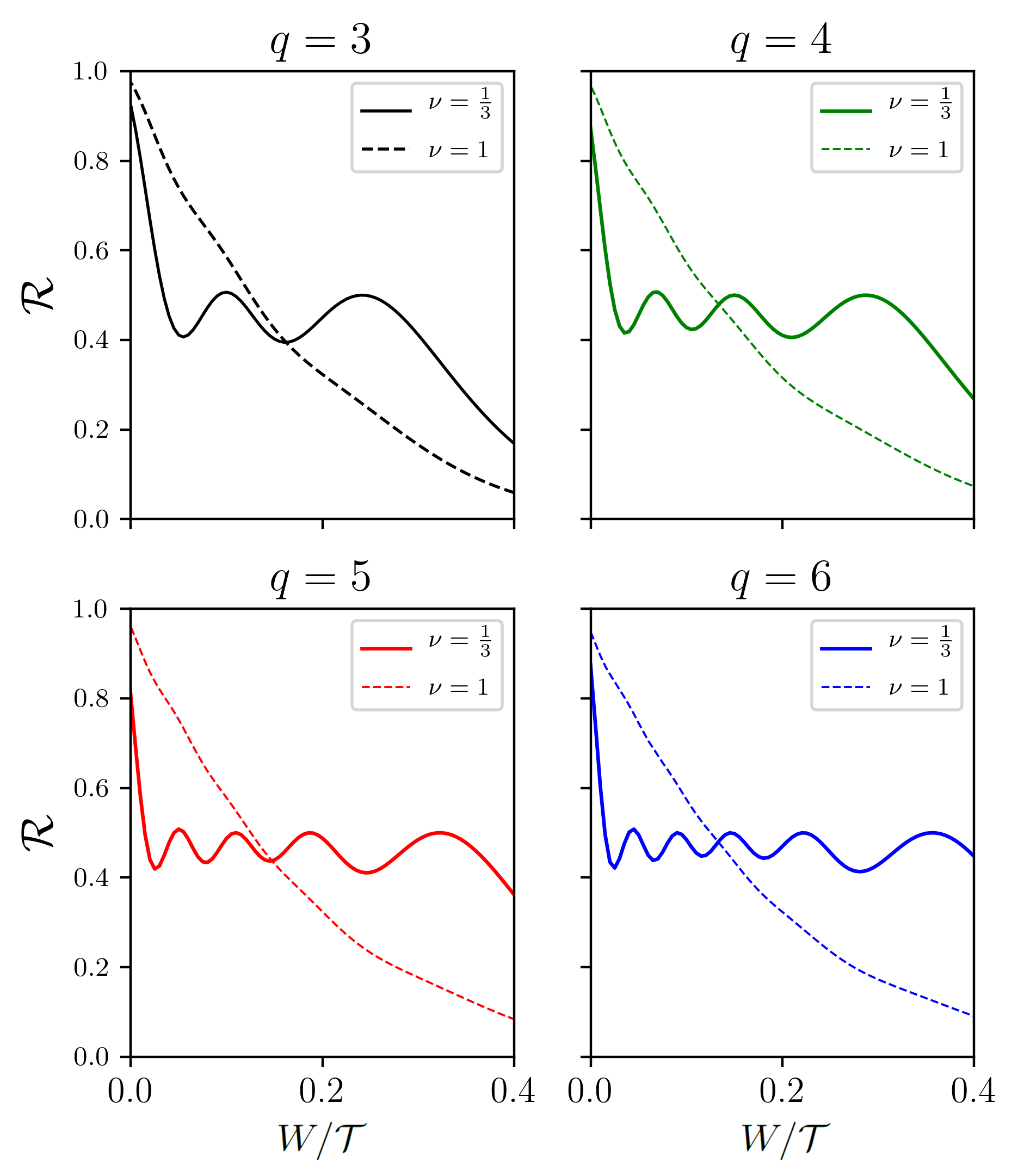}
	\caption{(Color online) Ratio $\mathcal{R}$ as a function of $ W/\mc T$ for $q=3$, $q=4$, $q=5$, $q=6$. The integer case (dashed lines) and the fractional case for $\nu=\frac{1}{3}$ (solid lines) are compared. The other parameters are $t_D=0.5 \mathcal{T}$ and $\omega=0.01 \omega_c$.}
	\label{ratioeta}
\end{figure}

By increasing the ratio between the width of the pulses and the period, the peak-to-valley amplitude of oscillations is enhanced for fractional filling factors, while for the integer case the situation is qualitatively unchanged, as depicted in Fig. \ref{fig:ratio2}. The principal downside is that some of the oscillations that are clearly visible for sharper pulses are now lost, since pulses belonging to neighboring periods start to overlap significantly.
Therefore, the choice of increasing the ratio $W/\mc T$ makes it easier to observe the presence of oscillations in the current-current correlators, even though some dips inevitably disappear.
Complementary information can be drawn by fixing a value of the delay $t_D$ and inspecting the shape of the ratio $\mathcal{R}$ as the ratio $W/\mc T$ is varied. The plots of $\mathcal{R}$ as a function of $W/\mc T$ for different values of $q$ are shown in Fig. \ref{ratioeta}, where we set $t_D=0.5\mathcal{T}$ since the signal is bigger and oscillations are more pronounced for such a value of the delay. Interestingly, the integer and the fractional cases show a dramatically different behavior. In the former case, the ratio is smoothly decreasing without any particular feature. In the latter, conversely, it oscillates for quite a large interval of $W/\mc T$, before eventually decreasing. Furthermore, the number of peaks appearing for fractional filling factors is exactly equal to $q$. This additional experimental investigation could significantly help in discriminating between the crystallized and the non-crystallized regime.
Finally, it is worth noting that the same behavior of the ratio can be observed for all filling factors in the Laughlin sequence. Such an universality tells us that interactions in Laughlin FQH states are always strong enough to induce a complete crystallization.

\section{Conclusions \label{sec:Conclusions}} The strongly correlated phase of FQH systems is able to crystallize Levitons, soliton-like excitations in the realm of condensed matter, after their tunneling at a QPC. This process rearranges the excess density of Levitons in a regular oscillating pattern, showing as many peaks as the number of injected particles. The amplitude of the oscillation gets enhanced by increasing the strength of interactions. The crystallization of Levitons represents an electronic counterpart of soliton crystals realized with photons in optical fiber setups. Experimental evidence of this effect can be found in a Hong-Ou-Mandel interferometer, where unexpected dips in the noise reveal the crystallization mechanism. This kind of experiment is within reach for nowadays technology. Possible extensions include the investigations of related setups as optimal sources for fractionally charged single-anyons \cite{Glattli16_pss}, as well as crystallization of Levitons in the exotic $5/2$ FQH state \cite{Willett87}.

\begin{acknowledgments}
We are grateful to E. Bocquillon, G. F\`eve and D. C. Glattli for useful discussions. L.V.\ acknowledges support from CNR SPIN through Seed project ``Electron quantum optics with quantized energy packets''. This work was granted access to the HPC resources of Aix-Marseille Universit\'e financed by the project Equip@Meso (Grant No. ANR-10-EQPX-29-01). It has been carried out in the framework of project ``1shot reloaded'' (Grant No. ANR-14-CE32-0017) and benefited from the support of the Labex ARCHIMEDE (Grant No. ANR-11-LABX-0033), all funded by the ``investissements d'avenir'' French Government program managed by the French National Research Agency (ANR). The project leading to this publication has received funding from Excellence Initiative of Aix-Marseille University - A*MIDEX, a French ``investissements d'avenir'' programme.
\end{acknowledgments}
\vspace{10mm}
\appendix
\section{Equations of motion in the presence of a voltage drive \label{app:equation}}
We consider the edge modes $\Phi_{R/L}$ in the presence of two generic voltages $\mc V_{R/L}(x,t)$ coupled separately with right and left propagating states respectively. The Hamiltonian reads 
\begin{align}
	H_0 + H_{\rm s}& = \sum_{r=R,L} \Big\{ \frac{v}{4\pi} \int dx \left[ \partial_x \Phi_r(x) \right]^2+  \nonumber\\&\pm \frac{e \sqrt{\nu}}{2\pi} \int dx \, \mc V_{r}(x,t) \partial_x \Phi_{r}(x) \Big\},
\end{align}
with the upper (lower) sign referring to the mode $R$ ($L$). Equations of motion for bosonic fields are
\begin{equation}
(\partial_t \pm v\partial_x) \Phi_{R/L}(x,t) = -e \sqrt \nu \mc V_{R/L}(x,t)
\end{equation}
and are solved by
\begin{equation}
	\label{eq:sol_eq_motion}
	\Phi_{R/L}(x,t) = \phi_{R/L}(t_\mp) - e \sqrt \nu \int_0^t ds \mc V_{R/L}[x \mp v(t-s),s],
\end{equation}
where $\phi_{R/L}(t_\mp)$ are the fields at equilibrium. Due to the linear dispertion relation of quantum Hall edge states in the Laughlin sequence, they evolve chirally and can be written as a function of one single variable $t_\mp = t \mp \frac x v$.
Using the factorization $\mc V_{R/L}(x,t) = \Theta(\mp x-d) V_{R/L}(t)$, which is reasonable in the case of two homogeneous, semi-infinite contacts driven with time-dependent pulses $V_{R/L}(t)$, we get
\begin{equation}
	\Phi_{R/L}(x,t) = \phi_{R/L}(t_\mp) - e \sqrt \nu \int_0^{t_\mp -\frac d v } ds V_{R/L}(s).
\end{equation}
The constant time shift $d/v$ has no physical effect in our calculations and we can safely neglect it. It's worth noticing that from the bosonization identity one has
\begin{equation}
	\psi_{R/L}(x,t) = \frac{\mc F_{R/L}}{\sqrt{2\pi a}}e^{-i \sqrt \nu \phi_{R/L}(t_\mp)} e^{i \nu e \int_0^{t_\mp} ds V_{R/L}(s)}.
\end{equation}
Thus, quasiparticle fields $\psi_{R/L}(x,t)$ experience a phase shift due to the presence of the oscillating voltage $V_{R/L}(t)$.
\bibliography{levitons_biblio_Flavio}

\begin{thebibliography}{57}
\bibitem [{pos()}]{position}%
  \BibitemOpen
  \href@noop {} {}\bibinfo {note} {Let us notice that the excess density
  backscattered into the lower, left-moving channel propagates rigidly towards
  left. Therefore, the temporal profile of $\Delta \rho_L(x,t)$ is identical
  for all positions $x<0$ (apart from a constant delay).}\BibitemShut {Stop}%
\bibitem [{del()}]{delay}%
  \BibitemOpen
  \href@noop {} {}\bibinfo {note} {Here the constant time shift $d/v$ due to
  the finite positions $x=\pm d$ at which the density operators are evaluated
  has been included into a redefinition of the integration variables $t$ and
  $t'$.}\BibitemShut {Stop}%
\end{thebibliography}

\end{document}